# Diffusion of Oligonucleotides from within Iron-Crosslinked Polyelectrolyte-Modified Alginate Beads: A Model System for Drug Release

Vladimir Privman,[a]* Sergii Domanskyi,[a] Roberto A. S. Luz,[b,c]

Nataliia Guz,[b] M. Lawrence Glasser,[a] Evgeny Katz[b]**

[a]Department of Physics, Clarkson University, Potsdam, NY 13676

[b]Department of Chemistry and Biomolecular Science, Clarkson University, Potsdam, NY 13676

[c]Instituto de Química de São Carlos, Universidade de São Paulo, São Carlos, SP 13560-970, Brazil

*privman@clarkson.edu, http://www.clarkson.edu/Privman, +1-315-268-3891

**ekatz@clarkson.edu, http://www.clarkson.edu/~ekatz, +1-315-268-4421

**ABTRACT:** We developed and experimentally verified an analytical model to describe diffusion of oligonucleotides from stable hydrogel beads. The synthesized alginate beads are $Fe^{3+}$-cross-linked as well as polyelectrolyte-doped for uniformity and stability at physiological pH. Data on diffusion of oligonucleotides from inside the beads provide physical insights into the volume nature of the immobilization of a fraction of oligonucleotides due to polyelectrolyte cross-linking, i.e., the absence of the surface-layer barrier in this case. Furthermore, our results suggest a *new simple approach to measuring the diffusion coefficient* of the mobile oligonucleotide molecules inside hydrogel. The considered alginate beads provide a model for a well-defined component in drug release systems and for the oligonucleotide-release transduction steps in drug-delivering and biocomputing applications. This is illustrated by destabilizing the beads with citrate that induces full oligonucleotide release with non-diffusional kinetics.

## 1. INTRODUCTION

Chemical systems utilizing stimuli-responsive materials and releasing preloaded molecules or nano-size species upon receiving external signals, have recently attracted interest due to their promise for various, mostly biomedical applications.[1-6] Designs involving different processes, of varying degree of complexity have been reported, with the release process activated by physical or chemical signals,[7] such as light,[8] magnetic field,[9] temperature change,[10] pH variation,[11,12] presence of redox-species,[13] small biomolecules (e.g., glucose)[14] or proteins,[15] and the presence of bacterial cells.[16] The majority of the systems developed were designed for biomedical applications; specifically, for signal-triggered drug release,[17-20] particularly in the framework of the emerging field of theranostics.[21] DNA release[22] triggered by various external signals (e.g., electrochemical,[23,24] photochemical,[25] thermal,[26] etc.) is of interest for many applications including gene delivery therapy[27] and biomolecular computing.[28] Polyelectrolyte micro- or nano-capsules loaded with encapsulated substances (including DNA) freed upon application of the triggering signal are particularly promising systems for controlled molecular-release.[29,30]

Alginate, a natural polymer with many unique properties, notably, biocompatibility,[31] has attracted attention for biomolecular encapsulation followed by slow leakage, unless fast controlled release is signal-triggered.[32,33] Because of alginate's ability to be ionically cross-linked with multivalent metal cations entrapping biomolecules into the biopolymer hydrogel matrix, numerous reports have been published on the encapsulation of proteins/enzymes,[34-38] DNA,[39,40] cells[41,42] and other (bio)molecular species, e.g., drugs,[43,44] in alginate gels with the retention of their full biological activity. In most of the systems for biomolecular encapsulation and release, the alginate polymer is cross-linked with $Ca^{2+}$ cations to yield a gel in the form of microcapsules capable of the slow release of entrapped molecules passively or in response to changed environmental conditions, through the controlled degradation of the assembly.[45] On-demand signal-triggered release processes from alginate gels, irrespective of the environmental conditions, are much less explored[46] and they are usually based on the use of $Fe^{3+}$-cross-linking cations.[7] For instance, electrochemical reduction of $Fe^{3+}$ to $Fe^{2+}$ cations within the alginate gel results in the gel dissolution and fast release of the entrapped species[47,48] (note that $Fe^{2+}$ cations are not capable of alginate cross-linking). In this case the alginate gel is organized as a thin film on an electrode surface. This process can be triggered by various biomolecular inputs[14-16,49] (small biomolecules, proteins, bacterial cells) or by combinations of them processed by biomolecular logic systems.[50,51] These inputs activate the bioelectrochemical process, ultimately resulting in the electrochemical reduction of $Fe^{3+}$ and alginate gel dissolution and molecular release.[7]

Alternatively, molecular release from $Fe^{3+}$-cross-linked alginate gel in the form of beads can be triggered by chemical reactions. The reactive species (e.g., citrate) produce more stable complexes with $Fe^{3+}$ cations, thus removing the latter from the alginate complex, destabilizing the gel and resulting in its dissolution and release of the pre-loaded biomolecules.[52,53] Citrate, triggering the release process, can be produced *in situ* by enzymatic reactions activated by biomarkers signaling certain biomedical dysfunctions, e.g., liver injury.[52] It should be noted that in some studies, in addition to the metal cross-linking, alginate gel was modified by adding other polymers/polyelectrolytes to produce composite structures.[54,55] These composite structures offer favorable environments for embedding certain biomolecular species or/and additional gel stabilizing. The added polymers were spread throughout the gel volume[40,43,44,46] or in some cases they were used as external coatings deposited layer-by-layer on the gel surface.[52,53,56,57] Depending on the specific composition of the alginate systems (e.g., the cross-linker used, polymer added, external solution composition, etc.) and the signal applied that triggers the release (e.g., citrate), the mechanism of the latter process and the fate of the alginate system can be different. In some cases, the alginate gel (e.g., in the form of beads) can be fully decomposed and dissolved,[52,53] obviously resulting in the release of all pre-loaded species. Under other circumstances the release of the embedded species could proceed without dissolution of the gel (this case will be considered in the present work).

While the major practical interest in many studies has been in the signal-triggered release,[7] the leakage of the loaded species from intact alginate beads or thin films, prior to the signal application, is also an important process. For practical realizations of effective signal-triggered release the non-controlled leakage should be minimized.[52] However, in other studies the slow leakage has been of interest and triggering signals were not used, thus focusing on the continuous slow delivery of the loaded species, e.g., drugs.[45] Therefore, studying both processes, slow leakage of the loaded species and their signal-triggered fast release, is important.

In the present work, our primary goal has been testing a detailed theoretical model of diffusional transport of oligonucleotide species (for brevity termed DNA in the following text) leaking from the *otherwise stable* alginate beads without any signal activating their release. For comparison, we also considered the release process triggered by the citrate input, which results in a significantly different process. The theoretical model was tested on the experimental data obtained for DNA release from alginate beads. The model, when applicable, can yield interesting physical insights into the mobility and transport of DNA in the hydrogel (particularly in alginate) matrix. Specifically, one- or two-parameter data fitting can provide information on the nature of the diffusion process inside the gel, on the fraction of DNA that is mobile, on the effects of the added stabilizing polyelectrolyte, and *yield the value of the diffusion coefficient*, $D$, of DNA in the hydrogel. The latter is actually a new simple method for a fast, straightforward experimental estimate of $D$.

Typically, description of the diffusional processes of biomolecules and polymers in cross-linked polymer networks is a challenging problem.[58-62] Methods for extracting the parameters of the diffusion process from experimental data in such systems are therefore of interest. Specifically, for hydrogels, the diffusion coefficient can depend on many physical, chemical, and network properties.[58,59] Herein, we demonstrate a new, relatively simple approach for directly extracting the value of the diffusion coefficient inside the hydrogel network from data on the time-dependent DNA release measured outside the gel in the surrounding solution. Other methods for measuring the diffusion coefficient of various (bio)molecules in hydrogels have been used in the past.[58,62-65] Our approach is not only complementary to earlier-developed methods, but also allows us to establish that some of the loaded DNA appears to be immobilized, whereas the rest can diffuse out of the beads into the solution. Furthermore, our modelling offers conclusions on the dramatic effects of added stabilizing polyelectrolyte on the release process and confirms that there are no special boundary effects on the mobile-fraction of the DNA as it diffuses out of the beads.

## 2. EXPERIMENTAL SECTION

### 2.1. Materials and instrumentation

Sodium alginate from brown algae (medium viscosity, ≥2000 cP), fluorescein isothiocyanate (FITC), citric acid, 3-(*N*-morpholino)propanesulfonic acid (MOPS-buffer), Triton X-100, dimethyl sulfoxide (DMSO) and other standard organic and inorganic materials and reactants were obtained from Sigma-Aldrich, J.T. Baker or Fisher Scientific and used without further purification. Custom made DNA oligonucleotide with a fluorescent label, 5' 6-FAM-TGC AGA CGT TGA AGG ATC CTC, was purchased from Integrated DNA technologies. Selection of this specific DNA sequence was motivated by its use in certain biocomputing systems,[28] though for the present study the sequence is not important and other DNA sequences of comparable length could be used. Poly(allylamine hydrochloride) (PAH) was purchased from Polysciences Inc. All experiments were carried out in ultrapure water (18.2 MΩ·cm; Barnstead NANOpure Diamond).

Fluorescent measurements of released DNA were performed using a fluorescent spectrophotometer (Varian, Cary Eclipse). Cross sections of the FITC-labeled-PAH/alginate beads were imaged using C1 Eclipse Nikon TE2000U fluorescent confocal microscope. Optical microscope was used for physical size measurements and monitoring citrate-stimulated bead reaction.

### 2.2. Experimental system and procedures

*Fabrication of DNA-loaded alginate hydrogel beads:* Alginate beads loaded with DNA were prepared following the standard procedure.[45] Sodium alginate was dissolved in water at 40 °C under stirring to yield a 1% (w/v) solution. Then 1.0 mL of alginate solution was mixed with 20 μL (0.1 mM) of DNA solution made in autoclaved water. The alginate–DNA solution was then sprayed through a 31-gauge needle forming droplets, the size of which was controlled by the rate of the plunger depression. On contact with the $FeCl_3$ (1% w/v) and PAH (0, 0.5, 0.75 or 1% w/v) dissolved in deionized water, the droplets, incubated in this cross-linking environment for 15 minutes, changed into alginate hydrogel spheres. The spheres were then transferred from the FeCl$_3$/PAH solution to a Petri dish where they were washed in water. This procedure can yield well-defined uniform-size spheres with average diameters in the range of 0.5 mm to 2 mm. Most of our experimental data were collected for beads with average diameter of 0.6, 0.9 and 1.8 mm.

*Preparation of PAH with a fluorescent label:* FITC, 4 mg, was dissolved in 0.5 mL of DMSO and protected from light. PAH, 500 mg, was dissolved in 6 mL of water (prior to the PAH dissolution the pH was adjusted to 8.4 using NaOH). Both solutions were mixed and kept at room temperature in dark for 2 days. To remove the excess of non-reacted FITC, a 10 kDa Nanosep® centrifugal device was used. The FITC-labeled PAH was used in experiments that imaged the PAH distribution in the alginate beads. The alginate beads containing FITC-labeled PAH were prepared in the same way as those using unlabeled PAH, but excluding DNA.

*Release of DNA from the alginate beads:* The DNA release was monitored by fluorescent spectroscopy at 518 nm (note that the DNA was labeled with a fluorescent dye, the excitation peak 485 nm). For these experiments 10 nearly identical beads were placed at the bottom of a cuvette containing aqueous solution with 2 mL of 50 mM MOPS-buffer, pH 7.4, 50 mM MgCl$_2$, 20 mM KCl, 120 mM NaCl, 0.03% Triton X-100, 1% DMSO with or without added citric acid; see later sections. Note that citric acid used as the signal triggering DNA release is termed "citrate" in what follows, because for our pH values it is dissociated. The solution was periodically mixed to provide uniform concentration of the released substance outside the beads, and then the fluorescence of the solution (well away from the beads) was measured every 10 minutes. The intensity of fluorescence at wavelength 518 nm provides a measure of the DNA release as a function of time. It should be noted that the composition of the solution for the release experiments, particularly the introduction of the Mg$^{2+}$ cations, was motivated by the expected use of the released DNA in a biocomputing system,[28] which is outside of the scope of the present study.

## 3. THEORETICAL SECTION

### 3.1. General considerations

In this work we do not attempt to fully describe the complicated dynamics of alginate beads' loss of cross-linking, potential breakup, and leakage of the loaded DNA in aqueous environments at physiological pH, caused by physical transport processes or chemical processes (e.g., application of citrate). Generally, such processes are complicated, the beads are in many cases non-uniform, and many parameters are needed to quantify the dynamics. Instead, we focus only on a model system that allows one to confirm and study a single dominant process: diffusion of the loaded DNA molecules out of the bead. For this purpose, the bead must be properly prepared: uniform and relatively stable, which occurs in the appropriate range of sizes, as described in the Experimental section and also in Results and Discussion.

We assume that the DNA released is uniformly distributed in the solution (which is periodically mixed). Its concentration is so much smaller than the concentration inside of the beads that it can be set to zero when diffusion within the beads is considered. Experiments (reported in the Results and Discussion section) revealed that there is no substantial surface layer of PAH polyelectrolyte formed around the beads when PAH is added during their preparation. Therefore, considering the mixing outside, we can assume zero boundary conditions (no reflection at the bead surface) for diffusion. Instead, the PAH polymer is distributed rather uniformly in the beads, producing the alginate/PAH composite structure and causing some of the DNA to be immobilized, as discussed later.

### 3.2. The diffusional model

The radial diffusion equation in three dimensions,

$$\frac{\partial c}{\partial t} = D \frac{1}{r^2} \frac{\partial}{\partial r} \left( r^2 \frac{\partial c}{\partial r} \right), \tag{1}$$

is assumed to hold inside a bead of radius $R$, where $D$ is the diffusion coefficient. The concentration, $c(r \leq R, t)$, of the DNA that can diffuse, i.e., is not immobilized, as discussed later, is assumed initially constant throughout the volume of the bead:

$$c(r \leq R, t = 0) = c_0 . \tag{2}$$

The solution can be obtained analytically by several approaches, as outlined, for instance, in Ref. 66, where earlier literature is cited. Here we find convenient to consider the Laplace Transform

$$C(r,s) \equiv \int_0^\infty dt\, e^{-st} c(r,t) . \tag{3}$$

The transformed equation

$$-c_0 + sC = D \frac{1}{r^2} \frac{\partial}{\partial r}\left(r^2 \frac{\partial C}{\partial r}\right) \tag{4}$$

should be solved for $r \leq R$, with the zero-concentration boundary condition

$$C(R, s) = 0 . \tag{5}$$

Equation (4) is then solved to get a finite $C(r = 0, s)$ as the second boundary condition. The solution is

$$C(r,s) = \frac{c_0}{s}\left(1 - \frac{R \sinh r\sqrt{s/D}}{r \sinh R\sqrt{s/D}}\right) . \tag{6}$$

We only need to consider the total amount of DNA left in a typical bead, because the experiments detect the DNA that leaks outside the beads. Thus, we define

$$n(t) = \int_0^R 4\pi r^2 dr\, c(r,t), \tag{7}$$

the Laplace transform of which then follows by integrating Eq. (6):

$$N(s) = \frac{4\pi c_0 R}{3}\left[\frac{R^2}{s} + \frac{3D}{s^2} - \frac{3R\sqrt{D}}{s^{3/2}}\coth\left(\sqrt{\frac{s}{D}}R\right)\right]. \tag{8}$$

By utilizing the Inverse Laplace Transform formula 5.9(26) on page 257 and also the second (from the top) definition on page 388 of Ref. 67, we obtain

$$n(t) = \frac{4\pi c_0 R^3}{3}\left[1 + \frac{3Dt}{R^2} - 3\int_0^{tD/R^2} du\, \frac{1}{\sqrt{\pi u}}\sum_{k=-\infty}^{k=\infty} e^{-k^2/u}\right], \tag{9}$$

where the sum under the integral is related to one of the standard Elliptic Functions (Theta Functions), see Supplementary Data.

The initial amount of DNA in the bead is

$$n_0 = 4\pi c_0 R^3/3. \tag{10}$$

We note that the total amount of the initially mobile DNA that leaks into the solution (from several nearly identical beads), detected in fluorescence measurements, is proportional to $n_0 - n(t)$. The time dependence of the fraction, $f(t)$, of the maximum amount of DNA, which leaked out at time $t$, then, provided the radii, $R$, of the beads are known, only depends on the diffusion coefficient $D$ of the DNA inside the beads:

$$f(t) = \frac{n_0 - n(t)}{n_0} = 3\int_0^{\frac{tD}{R^2}} du\left[\left(\frac{1}{\sqrt{\pi u}}\sum_{k=-\infty}^{k=\infty} e^{-\frac{k^2}{u}}\right) - 1\right] \tag{11}$$

$$= 1 - 6\pi^{-2} \sum_{j=1}^{j=\infty} j^{-2} e^{-\pi^2 j^2 tD/R^2} ,$$

where the second expression is derived in the Supplementary Data.

The expressions in Eq. (9) or (11) are exact, and in fact variants of them are available in the literature, e.g., Sec. 6.3.1 of Ref 66. However, they are not convenient for numerical evaluation in data fitting. Instead, considering the typical data to be modelled, it should suffice to use good-precision short-time and large-time approximations. Let us introduce the dimensionless time variable

$$\tau = Dt/R^2 . \tag{12}$$

In our data fitting, we used

$$f(\tau) \simeq 6\pi^{-1/2} \tau^{1/2} (1 + 2e^{-1/\tau}) - 12 \, \text{Erfc}\,(\tau^{-1/2}) - 3\tau, \quad \text{for } \tau < 0.40 ,$$

$$f(\tau) \simeq 1 - 6\pi^{-2} e^{-\pi^2 \tau} , \quad \text{for } \tau > 0.40 , \tag{13}$$

where the first, short-time, approximation was obtained by retaining the leading ($k = 0$ and $\pm 1$) contributions to the sum in the integrand in Eq. (11) for small $u$,

$$\sum_{k=-\infty}^{k=\infty} e^{-k^2/u} \simeq 1 + 2e^{-1/u} , \tag{14}$$

and integrating. The large-time expression is obtained by keeping only the $j = 1$ term in the sum in the last expression in Eq. (11). The two expressions in Eq. (13) have comparable accuracy at $\tau \simeq 0.40$. The accuracy of this approximation is about 0.05%, much smaller than typical experimental uncertainties. All our data fitting was for values of $\tau$ not exceeding 0.70.

For data fitting programs that do not include the Complementary Error Function, Erfc, a very accurate approximation (0.0025%) for any positive value of $\tau$, derived from Ref. 68, p.299, 7.1.25, can be used:

$$\text{Erfc}(\tau^{-1/2}) \simeq \frac{\tau^{3/2} + 0.2823613\,\tau + 0.0770324\,\tau^{1/2}}{(\tau^{1/2} + 0.47047)^3} e^{-1/\tau}. \tag{15}$$

## 4. RESULTS AND DISCUSSION

### 4.1. Stability of the $Fe^{3+}$-Cross-Linked Alginate Beads

Generally, metal-ion cross-linked alginate beads can release loaded materials (here DNA) by means of several processes, in many cases ongoing simultaneously.[53,57,59,69] In addition to DNA diffusing out of the beads, the beads themselves can swell or break up in aqueous solutions, either on their own[45] or by the destabilization of a protective surface layer.[52,53,57] These processes are usually induced or significantly accelerated by adding chemicals that reduce the number of the cross-linking cations or re-complexate them. The latter is the case[52,53] when $Fe^{3+}$-cross-linked alginate reacts with citrate. Generally, $Fe^{3+}$-cross-linking can be reduced by chemical or electrochemical means.[7] However, polyelectrolyte dopants, such as PAH here, introduced during the beads' synthesis can have the opposite effect of stabilizing the beads as well as decreasing the mobility of the loaded material.[40,43,44,46] This stabilizing effect is particularly pronounced when the negatively charged alginate polymer forms a complex with a positively charged polyelectrolyte (e.g., polyethyleneimine, polylysine, or polyallylamine; the latter used in the present study).

In this work, we devised a synthesis procedure yielding alginate beads that remain stable at physiological pH. With enough PAH added, even the presence of citrate does not cause the beads to swell or break up, but only results in the change of their light-brown/yellow color to become more transparent, as illustrated in Figure 1, which is indicative of the loss of iron ions. A more detailed illustration showing color image sequences with and without added citrate is provided in the Supplementary Data, Figure S1. This indicates that $Fe^{3+}$ cations are replaced and as a result internal cross-linking is decreased, while the size and shape of the beads are preserved due to the alginate/PAH-complex formation. Addition of citrate also causes the originally immobilized DNA fraction to become mobile, and all the DNA leaks out of the beads. This will be illustrated in a later subsection.

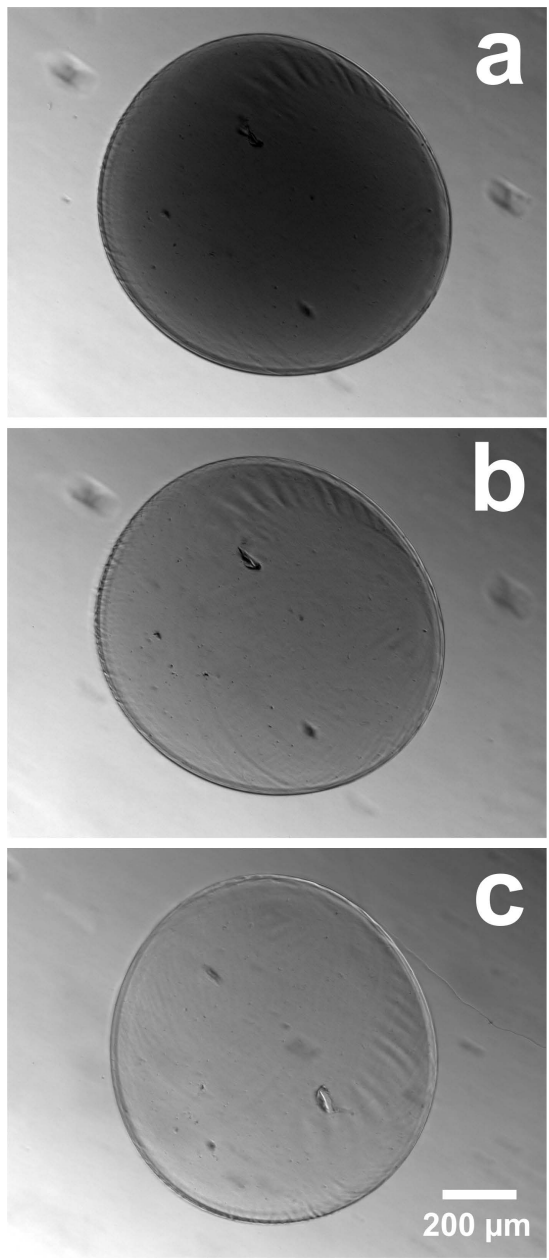

**Figure 1.** An illustration of a 0.9 mm bead prepared with PAH. When the beads were prepared with enough PAH, here 1% (w/v), no swelling or breakup was observed. Panel (a) shows the original freshly prepared bead, while panels (b) and (c) show the bead after reacting with 1 mM citrate, for 4 h and 20 h, respectively. Note the transparency change in the bead upon reaction with citrate. (Figure S1 in the Supporting Information offers additional details, by showing color image snapshots for a larger bead, with and without added citrate, for times up to 3 h.)

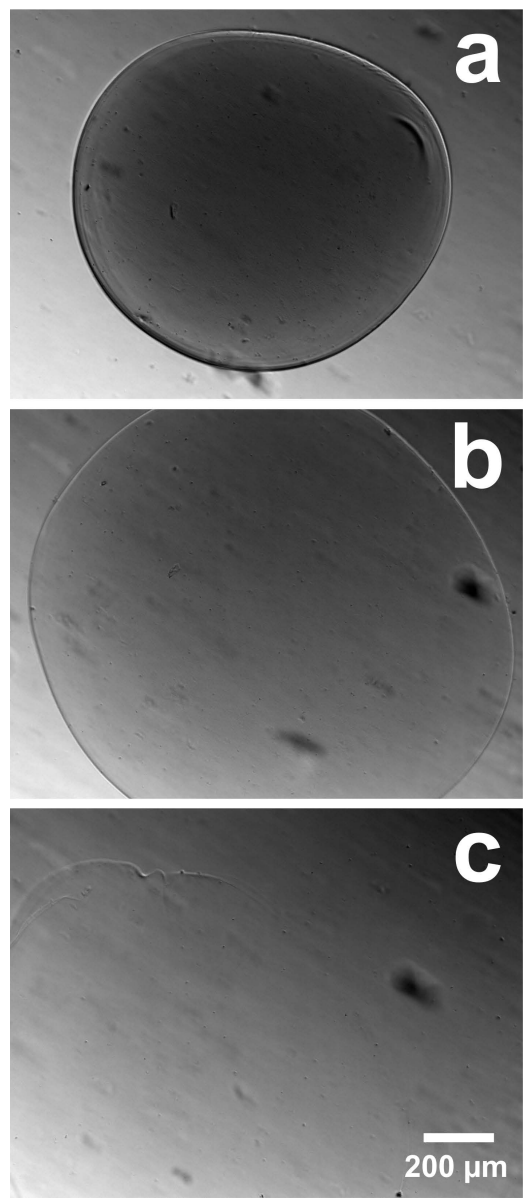

**Figure 2.** When the same beads as in Figure 1 are prepared without PAH, after addition of 1 mM citrate they swell and ultimately completely dissolve. Panel (a) shows the original freshly prepared bead, while panels (b) and (c) show the bead after reacting with 1 mM citrate, for 20 min and 50 min, respectively.

The role of PAH for our beads seems to be volume-wide stabilization. Figure 2 illustrates this; when the same beads (as in Figure 1) are synthesized without the added PAH, exposure to citrate causes them to swell, eventually leading to their complete breakup/dissolution. The stabilizing role of PAH was further explored by fluorescent labeling of PAH (without DNA added). As shown in Figure 3, PAH is relatively uniformly distributed over the bead volume, rather than forming a surface layer. This "bulk" rather than "interfacial" effect of PAH is further confirmed by our modeling, as discussed later.

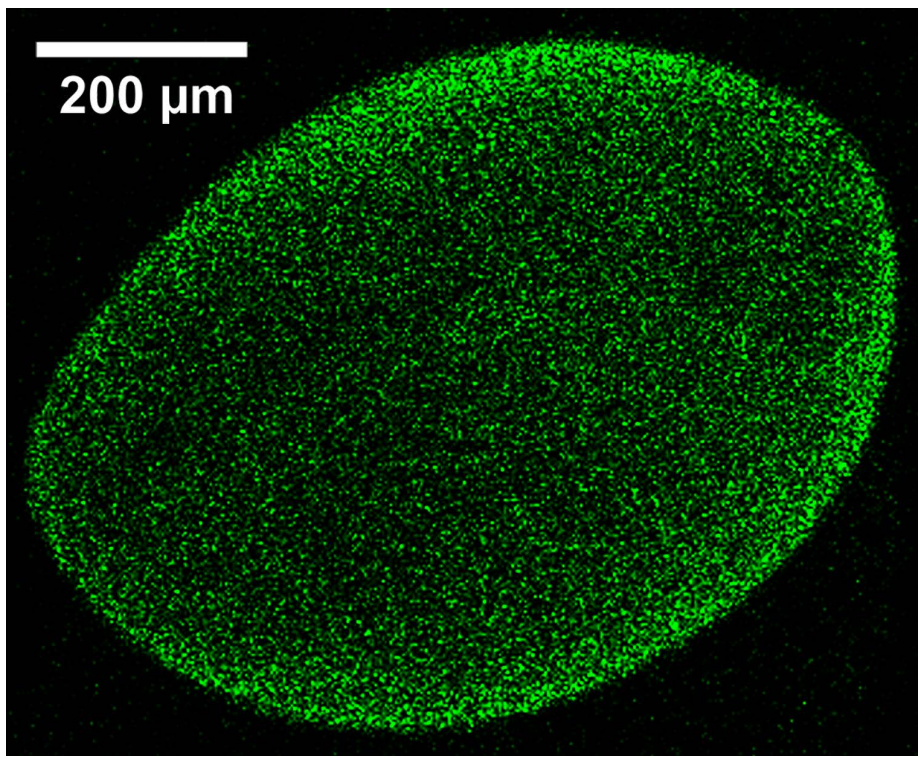

**Figure 3.** Fluorescent image of a bead cross-section (central one-third slice) obtained by confocal optical microscope. The beads were cut for imaging, which caused the observed distortion from circular cross-section. Such images were observed for beads of varying diameters. Here we show a 0.9 mm bead prepared with fluorescently labeled PAH, 1% (w/v), but without DNA. As seen from this illustrative image, PAH is distributed throughout the volume, rather than forming a surface layer.

**4.2. Verification of the Diffusional Model**

Detailed data were obtained and fitted for bead sizes $0.6 \pm 0.1$ mm, $0.9 \pm 0.1$ mm and $1.8 \pm 0.1$ mm, in order to test the proposed diffusion model detailed in the preceding section. Recall that the goal of this work has been to identify a synthesis procedure yielding beads from which DNA diffuses in a simple, single parameter controlled process, that parameter being the DNA diffusion coefficient, $D$, inside the beads. Beads that are too small are difficult to produce with the well-defined properties and uniform sizes required for our purposes. For beads of average sizes 0.6 and 0.9 mm, the experimental data are shown in the top and middle panels of Figure 4, for various amounts of PAH added during the synthesis. It is obvious from the data that addition of PAH immobilizes some of the DNA, which is an expected result of the charge interaction between negatively charged DNA and positively charged PAH. The maximum amount of DNA leaking out at large times decreases for larger amounts of the PAH initially added. An illustration of this is that for large enough quantity of PAH practically no release can be observed, as is seen in data set (d) in the middle panel of Figure 4.

After subtracting the background (seen in the data), the maximum quantity of DNA that leaks out of beads of average sizes 0.6 and 0.9 mm could be determined by using the largest-time value. Single-parameter fits of the values of $D$ were carried out according to the equations derived in the Theoretical section. The quality of the fits is rather good (see Figure 4, the top and middle panels). The results for the diffusion coefficient, $D$, will be discussed in next subsection. This not only validates the model for diffusion as the only dominant process of DNA leakage from within these beads, but also confirms that there are no special surface-layer effects that would require modification of the simple boundary condition assumed at the bead surface.

Beads of larger sizes are less well described by the uniform diffusion-only model, as can be seen in the bottom panel of Figure 4. This is likely attributable to the fact that larger beads are not fully internally uniform, despite the stability of the beads prepared with our procedure. This non-uniformity is a well-known property of all alginate-based hydrogel beads.[70] The data did not show full release of the mobile DNA for the large time considered (note that the time scales increase with the bead sizes, between the panels of Figure 4). The best least-square fits based on the single-parameter model are at best semi-quantitative to qualitative in this case. If instead of using the largest-time measured value for the maximum amount of DNA released, we make this quantity an adjustable parameter (thus, the fit becomes two-parameter), the quality of the model fits for the bottom-panel data (for large beads) in Figure 4 is only marginally improved, and the results for the diffusion coefficient, which are presented and discussed later, are not significantly changed. We remark that two-parameter fits were also done for smaller beads, for data in the top and middle panels in Figure 4, but no noteworthy changes in the fitted curves were observed.

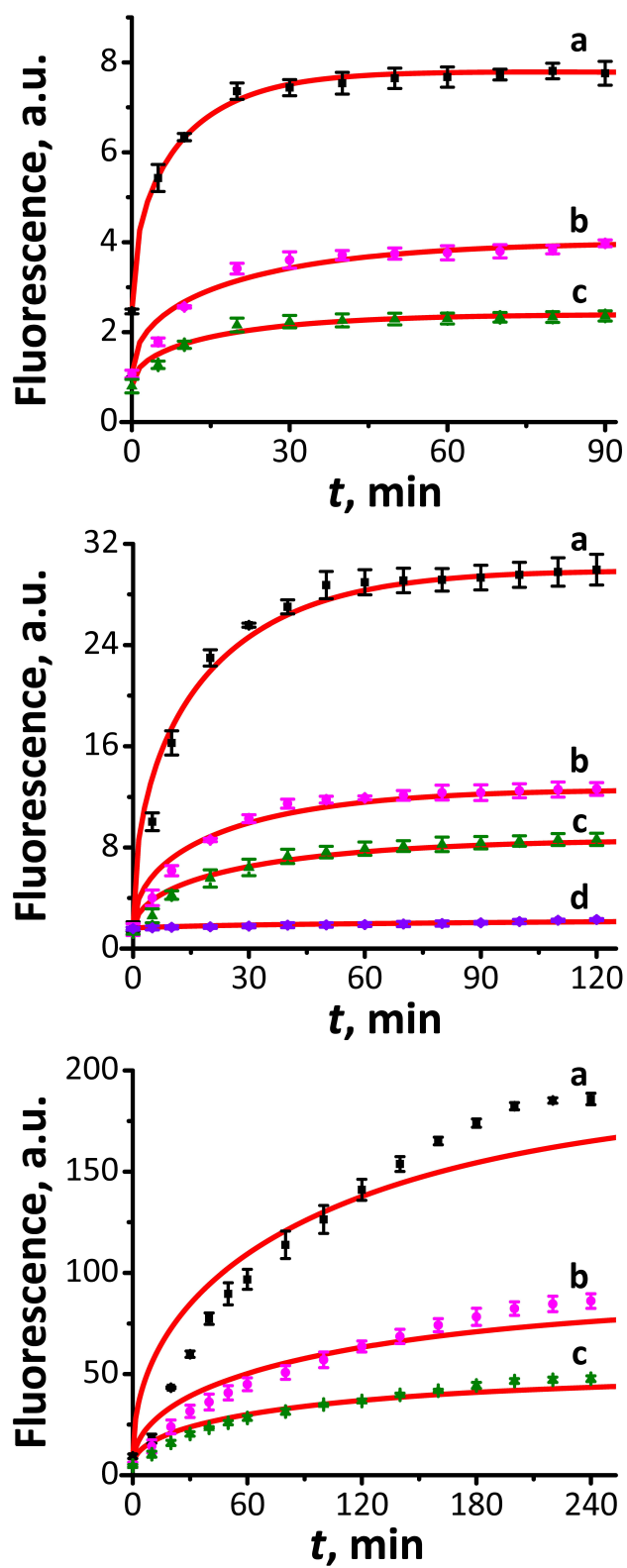

**Figure 4.** DNA leakage from the beads measured as fluorescence intensity at wavelength 518 nm, in arbitrary (instrumental readout) units (a.u.). Top panel: DNA-loaded beads have diameter $0.6 \pm 0.1$ mm; middle panel: $0.9 \pm 0.1$ mm; bottom panel: $1.8 \pm 0.1$ mm. Solid lines represent the model fit of the corresponding experimental data (symbols). For set (a) no PAH was added during the preparation stage, for sets (b), (c) and (d) PAH was added in the amounts of 0.5% (w/v), 0.75% (w/v) and 1% (w/v), respectively.

Interestingly, the largest quantities of DNA released, measured at the largest time or, when relevant, obtained by the two-parameter fitting just mentioned, were found to be rather precisely in proportion $0.6^3 : 0.9^3 : 1.8^3$, for three different bead diameters (0.6, 0.9, 1.8 mm), for fixed PAH concentration, 0.0% (w/v), 0.5% (w/v), and 0.75% (w/v). This confirms our assumption that the fraction of the DNA that is mobile to diffuse inside the beads after their synthesis, is initially uniformly distributed throughout the bead volume, and its density is approximately the same for different-size beads, determined only by the amount of PAH.

**4.3. Effects of the Degree of Alginate Doping with PAH on Diffusional Kinetics**

The fitted diffusion coefficients for various amounts of PAH used in the initial synthesis (but without any citrate added during the DNA leakage stage) are summarized in Table 1. The values obtained for $D$ are in agreement with earlier reported data for diffusion of similar-size relatively short biomolecules in hydrogels, tabulated in a comprehensive review.[59] For longer DNA chains, with over $M \approx 600$ base pairs, in a different hydrogel, diffusing out of cylindrical rather than spherical samples, the diffusion coefficients were found to be of order of $10^{-3}$ times smaller,[71] which, considering the ratio of the base-pair counts in the chains, denoted $M$, is consistent with the expectations of the reptation[72] theory of diffusional transport of chains in networks, that predicts that the diffusion constant in three dimensions is proportional to $M^{2\nu-3}$. In fact, if we take the numbers literally, then our data suggest that the ideal random-walk value for the correlation length critical exponent, $\nu = 1/2$, is more appropriate than the self-avoiding walk value, $\nu \approx 3/5$. However, the latter conclusion, while consistent with other such findings,[72,73] is in our case at best semi-quantitative because the compared systems differ in many other parameters.

Note that within the expected accuracy of ±10%, given the quality of the experimental data, the diffusion coefficients are consistent for beads of the three sizes considered, and seem to be determined by the amount of the PAH initially added in the solution during the beads' synthesis. Large quantities of PAH decrease the diffusion coefficient so that ultimately there is practically no leakage from the beads.

**Table 1.** Summary of the results for the fitted values of the diffusion coefficient, in units of $10^{-6}\,\text{cm}^2/\text{s}$, for the beads of different sizes, prepared with several concentrations of the doping PAH polyelectrolyte in the solution into which alginate was injected during the beads' preparation.

| Diameter (mm) | 0% (w/v) PAH | 0.5% (w/v) PAH | 0.75% (w/v) PAH | 1% (w/v) PAH |
|---|---|---|---|---|
| 0.6 ± 0.1 | 0.49 | 0.29 | 0.28 | |
| 0.9 ± 0.1 | 0.54 | 0.43* | 0.34 | 0.09*† |
| 1.8 ± 0.1 | 0.38† | 0.36 | 0.41 | |

---

*Two cases for which effects of the application of citrate were illustrated (see Figure 5).

†These two values for $D$ are not precise (cf. Figure 4).

In addition to decreasing the value of $D$ for the mobile fraction of the loaded DNA, PAH seems to cause a fraction of the DNA to be fully immobile. This can be seen by the differences in the saturation values, decreasing in the order of data sets (a), (b), (c), (d) in the panels of Figure 4. This observation was confirmed by illustrative experiments whereby citrate was added 60 min after the start of a typical "leakage" experiment, for beads of diameter 0.9 mm prepared with 0.5% (w/v) PAH, and for a case when there was practically no leakage, i.e., beads prepared with 1% (w/v) PAH; see Figure 5.

It is noteworthy that the maximum amount of DNA released, with citrate added reaches approximately the same values for both experiments shown and these values are also consistent with the amount of DNA released after large enough time when no PAH was used and citrate was not needed; cf. data set (a) in the middle panel of Figure 4. We attempted to fit the data for the access DNA released due to the effects of citrate, see Figure 5, for times $t \geq 60$ min, with the simple diffusional model used for leakage without citrate. However, both the single- and two-parameter fits clearly fail.

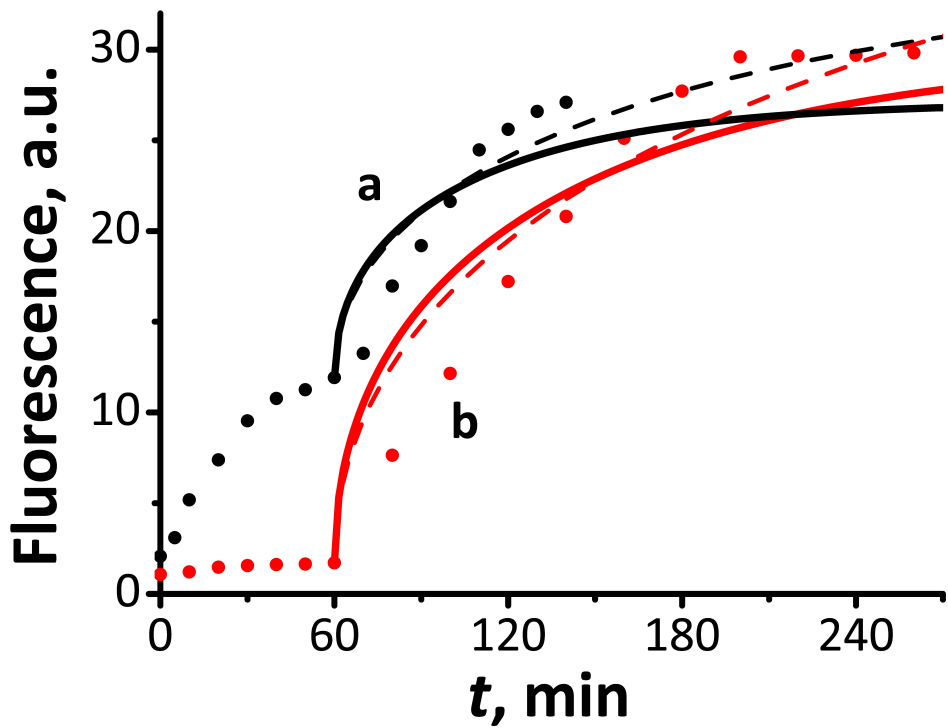

**Figure 5.** DNA leakage measured as fluorescence intensity at wavelength 518 nm for the beads of diameter 0.9 ± 0.1 mm, (a) for PAH added in the amount of 0.5% (w/v), and (b) for 1% (w/v) PAH. Citrate, 1 mM, was added to the solution after 60 min, causing fast release of the remaining DNA. The solid lines represent the best model 1-parameter fit assuming the simple diffusional model for the access DNA (on top of what would be released without the added citrate) for $t \geq 60$ min. The broken lines represent the best model 2-parameter fit (the saturation concentration taken as the second adjustable parameter).

These observations confirm the expectation that a fraction of DNA is trapped (immobile) in the beads due to effects of PAH. Once citrate is added, this fraction of DNA becomes mobile and is released as the beads' iron-ion cross-linking is degraded. However, the kinetics of this process is more complicated than the simple diffusional model. Its modeling would require consideration of the mechanisms of citrate penetrating the structure and decreasing cross-linking, as well as negating the effects of PAH, likely causing the effective diffusion coefficient of this fraction of the DNA to increase from zero as a function of time. This process will also introduce dependence of $D$ on the radial distance from the centers of the beads. Such modeling was not attempted here, because the leakage data on their own are not definitive enough to decide on the specific processes to be considered.

## 5. SUMMARY

Experimental protocol was developed for synthesis of relatively stable, at physiological conditions, $Fe^{3+}$-crosslinked and PAH-doped alginate beads of millimeter to submillimeter diameter. We established an approach to verify the purely diffusional nature of and to quantitatively characterize the process of DNA fragments' release from the otherwise stable alginate beads. The analytical model for transport of DNA from within the beads offers a new and relatively simple method to extract the diffusion coefficient of these molecules inside the hydrogel.

Our successful verification of the proposed model also confirms that no special surface boundary condition is necessary for non-zero concentration of the added PAH polyelectrolyte, indicating that the effect of this polyelectrolyte is internal to the beads in this case: it was found to immobilize a fraction of the DNA and it generally reduced the DNA's diffusion coefficient. The diffusion-only model is no longer valid for release kinetics in the case of beads destabilized with citrate. The quantitative study of the kinetic processes in the latter case[52,53] was outside the scope of the present work, which focused on DNA diffusion inside the hydrogel and extraction of the diffusion coefficient.

The experimental results and theoretical model are important for the general understanding of biomolecular leakage and release of biomolecules from hydrogels and particularly for DNA release from alginate beads. The latter process is important for various biomedical applications (e.g., drug delivery) and for activating DNA biocomputing[28] systems.

**ACKNOWLEDGEMENTS**

We are thankful to B. Fratto, Dr. M. Gamella, Prof. D. Kolpashchikov, and Prof. A. Melman for useful discussions and helpful suggestions. We gratefully acknowledge funding of our research by the NSF, via award CBET-1403208. Dr. R. A. S. Luz acknowledges CNPq (Post-Doctoral scholarship PDE processing number: 249152/2013-4) for support.

**Supporting Information** for this article is available on the WWW under http://dx.doi.org/10.1002/cphc.201501186.

# Supporting Information

*Derivation of the Expression Leading to the Large-Time Approximation*

Here we offer a concise derivation of the last expression in Eq. (11), useful in getting the large-time approximation given as the second of Eq. (13). We note that the sum in the integrand of Eq. (11) can be identified as one of the standard Theta Functions; see Ref. S1, page 877, relation 8.180.4:

$$\vartheta_3(0|i\pi x) \equiv \sum_{k=-\infty}^{k=\infty} e^{-\pi^2 k^2 x} . \qquad (A1)$$

With $1/\pi^2 u$ as the argument $x$ in Eq. (A1), the expression involving the integral in Eq. (11) can be written as

$$f(\tau) = 3 \int_0^\tau du \left[ \pi^{-1/2} u^{-1/2} \vartheta_3(0|i\pi^{-1} u^{-1}) - 1 \right] . \qquad (A2)$$

Next, we utilize the following Jacobi Imaginary Transformation[S2] for the Theta Function in Eq. (A2),

$$\vartheta_3(0|i\pi^{-1} u^{-1}) = \pi^{1/2} u^{1/2} \vartheta_3(0|i\pi u) , \qquad (A3)$$

yielding

$$f(\tau) = 3 \int_0^\tau du \left[ \vartheta_3(0|i\pi u) - 1 \right] = 3 \int_0^\tau du \left[ \left( \sum_{j=-\infty}^{j=\infty} e^{-\pi^2 j^2 u} \right) - 1 \right] , \qquad (A4)$$

where we again used Eq. (A1). Integration then yields the last expression in Eq. (11).

– S1 –

*Illustration of the Effect of Citrate on Stable Beads*

Here we illustrate the process of the loss of $Fe^{3+}$ cations cross-linking for beads that are prepared otherwise stable (not swelling or breaking apart) when exposed to citrate. Figure S1 (on the next page) shows snapshots of a bead with and without citrate added.

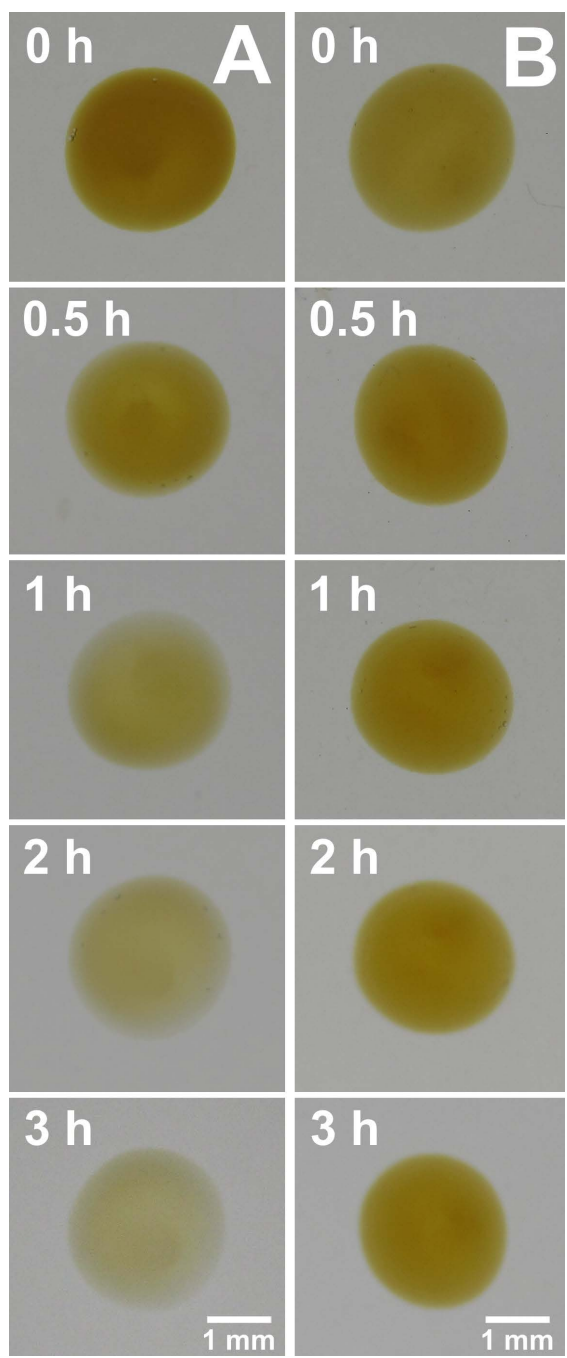

**Figure S1.** An illustration of the bead stability and the effect of added citrate, 1 mM, for a 2 mm bead prepared with 1% (w/v) PAH, cf. Figure 1. (A) The left sequence of images shows the discoloration (loss of $Fe^{3+}$ cations cross-linking), but no swelling caused by added citrate. (B) The right sequence shows that a bead remains stable when no citrate is added.